\documentclass[conference]{IEEEtran}

\IEEEoverridecommandlockouts

\usepackage[T1]{fontenc}
\usepackage[utf8]{inputenc}
\usepackage{cite}
\usepackage{amsmath,amssymb,amsfonts}
\usepackage{graphicx}
\usepackage{booktabs}
\usepackage{multirow}
\usepackage{array}
\usepackage{siunitx}
\usepackage{xcolor}
\usepackage{xurl}
\usepackage[hidelinks]{hyperref}
\usepackage{ifthen}
\usepackage{balance}
\usepackage{soul}

\newcommand{\name}{\textit{DC-CLM }}
\newcommand{\namenospace}{\textit{DC-CLM}}

\begin{document}

\title{\namenospace: Extending the WECC Composite Load Model for AI Data Center Dynamics}

\author{
 \IEEEauthorblockN{Md Kibria Saroare, Md Rubel Ahmed, Arif Hussain}
  Louisiana Tech University
 \IEEEauthorblockA{\{msa078,mahmed,ahussain\}@latech.edu}
 }
\maketitle

\begin{abstract}
The rapid growth of AI-driven data centers is introducing load behaviors that are not explicitly represented in conventional composite load models. This paper presents \namenospace, a workload-aware extension of the WECC composite load model that incorporates UPS-supported IT demand, mixed motor/VFD cooling loads, auxiliary demand, and training, inference, and idle workload profiles. A rule-based supervisory state machine represents grid, battery, and diesel operating states and captures temporary IT-load isolation and workload-dependent restoration following voltage recovery. The model is implemented in MATLAB/Simulink using positive-sequence phasor-domain simulation and evaluated under fault-induced delayed voltage recovery (FIDVR) conditions. For a study system with a 100~MW conventional composite load and a 200~MW data-center load, the long-duration voltage recovery is governed mainly by conventional stalled-motor thermal dynamics, while the first 100~ms after fault clearance is strongly workload-dependent. Training and inference produce approximately 50--52~mpu voltage variation and 751--755~MW active-power variation, compared with approximately 36~mpu and 654~MW for idle operation. The results demonstrate the value of incorporating data-center-specific load dynamics into transmission-level stability studies while highlighting the need for future measurement-based validation and more detailed converter-level modeling.
\end{abstract}


\section{Introduction}
Power-system load modeling has long been recognized as critical for accurate dynamic simulation and stability assessment\cite{Kundur1994,Kosterev2006WECC,Kosterev2008WECCProgress}. The WECC composite load model (CLM), developed to capture the aggregate behavior of diverse end-use devices, remains a standard representation for phasor-domain transient-stability studies\cite{Kosterev2008WECCProgress,Huang2019WECCComposite}. A key contribution of the WECC CLM is its ability to represent fault-induced delayed voltage recovery (FIDVR) caused by stalling of single-phase air-conditioning compressor motors\cite{Lesieutre2008ACMotor}. The model structure includes three-phase induction motors, a single-phase performance-based motor model, electronic loads, and static ZIP loads connected through a distribution equivalent network\cite{Huang2019WECCComposite,Ma2020WECCMath}. However, as power systems increasingly host AI-driven data centers and other emerging large loads, the WECC CLM requires further development to represent behaviors such as UPS ride-through, rapid load rebound, and workload-dependent demand variation \cite{NERC2026LargeLoads,ESIG2026LargeLoadsModeling}.
 
Bulk power-system planning and operation are being increasingly affected by the rapid expansion of data centers, whose large and fast-varying demand is changing load-growth patterns and challenging existing dynamic-modeling practices \cite{Shehabi2024USDataCenter,NERC2026LargeLoads,ESIG2026LargeLoadsModeling}. Industry studies project that U.S. data-center electricity demand could account for a substantial fraction of national load growth during the coming decade\cite{Shehabi2024USDataCenter,IEA2024Electricity}. Unlike conventional industrial loads, modern AI data centers can exhibit highly variable demand that contributes to faster and more complex grid dynamics, including rapid fluctuations during synchronized GPU training phases and bursty inference workloads \cite{Patel2024LLM,GinzburgGanz2026Survey}. Recent assessments have highlighted several gaps relevant to this work, including interconnection and reliability challenges associated with rapidly growing data-center demand \cite{NERC2026LargeLoads}, the need for improved phasor-domain representations of emerging large loads \cite{ESIG2026LargeLoadsModeling}, and the absence of conventional aggregate load models that capture UPS-mediated and converter-dominated data-center behavior \cite{Sun2022DCEmulator}. 

Despite the growing importance of these loads, conventional composite-load representations do not explicitly capture several behaviors relevant to large data centers. In particular, a conventional static or voltage-dependent electronic-load representation does not describe an IT load that is temporarily isolated from the grid through UPS action and subsequently restored according to supervisory transfer logic. It also does not explicitly distinguish between stall-sensitive motor-driven cooling equipment and VFD-driven cooling equipment or incorporate workload-dependent IT demand. Recent studies have investigated data-center power-distribution dynamics, converter-based representations, and transient-stability-oriented data-center models~\cite{Sun2022DCEmulator,JimenezRuiz2025DCModel}. At the same time, recent industry assessments have emphasized the need for improved dynamic representations of emerging large loads in system-level studies~\cite{NERC2026LargeLoads,ESIG2026LargeLoadsModeling}.

This work focuses on incorporating these data-center-specific behaviors into the established WECC composite load modeling framework. The proposed \name provides an aggregate system-level representation for examining how UPS-mediated IT-load isolation and restoration, mixed motor/VFD cooling behavior, and workload-dependent demand interact with conventional composite-load dynamics during voltage disturbances.
\begin{figure*}[!t]
    \centering
\includegraphics[width=\linewidth]{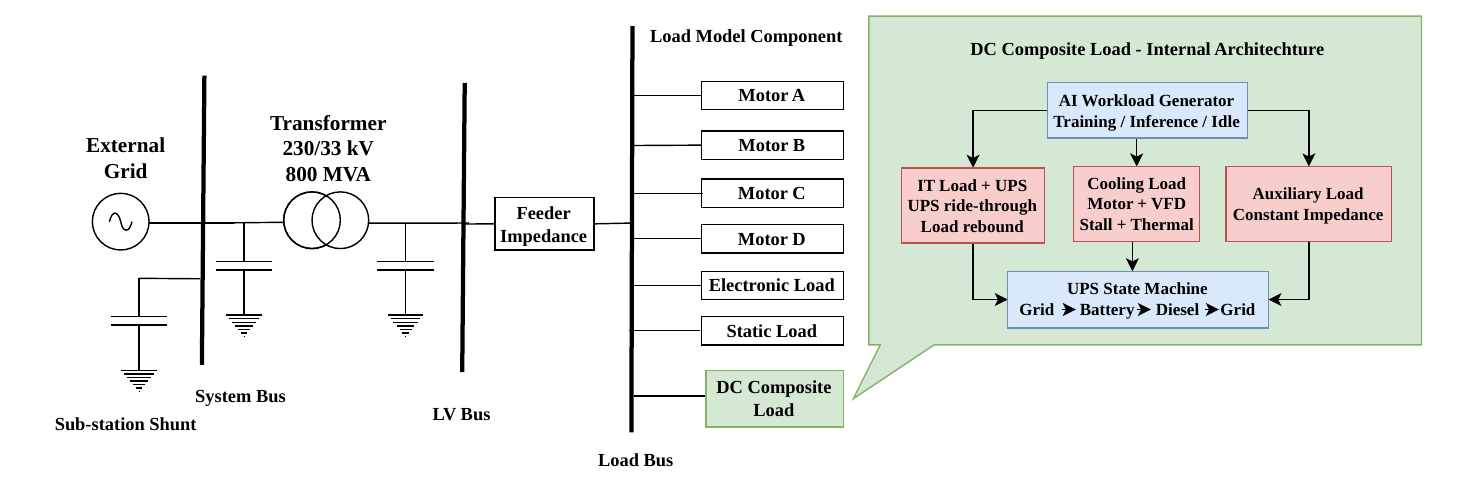}
    \caption{Architecture of the proposed \name integrated with the WECC composite load model~\cite{Huang2019WECCComposite}. The data-center component combines a workload-dependent IT load, UPS supervisory logic, mixed motor/VFD cooling demand, and auxiliary demand. The UPS state machine determines the grid-side availability of the IT load during and after a voltage disturbance.
    \vspace{-15pt}}
    \label{fig:architecture}
\end{figure*}

The main contributions of this work are as follows:
\begin{enumerate}
    \item A workload-aware data-center component is integrated with the WECC CLM, combining UPS-supported IT demand, mixed motor/VFD cooling, auxiliary load, and AI workload profiles.
    
    \item A rule-based supervisory model represents grid, battery, and diesel operating states, including temporary IT-load isolation and workload-dependent restoration.
    
    \item The proposed \name is implemented in positive-sequence phasor mode and evaluated under FIDVR conditions to distinguish long-duration voltage recovery from workload-dependent post-fault transients.
\end{enumerate}
The remainder of this paper is organized as follows. Section~II reviews the WECC composite load model. Section~III presents the proposed \name framework. Section~IV describes the positive-sequence phasor-domain implementation and simulation setup. Section~V presents the FIDVR case studies. Section~VI concludes the paper and discusses future research directions.

\section{WECC Composite Load Model}
\subsection{Model Structure}

The WECC CLM structure, shown in the left portion of Fig.~\ref{fig:architecture}, represents the aggregate load at a transmission bus through a distribution equivalent network and six load components\cite{Kosterev2008WECCProgress}. The equivalent network includes a substation transformer with on-load tap changer, a shunt capacitor, and a feeder represented by a $\pi$-section with impedance $R_{\mathrm{fdr}} + jX_{\mathrm{fdr}}$. At the load bus, the total active power is distributed among the component fractions as
\begin{equation}
P = \left(F_{\mathrm{ma}} + F_{\mathrm{mb}} + F_{\mathrm{mc}} + F_{\mathrm{md}} + F_{\mathrm{el}} + F_{\mathrm{st}}\right) P_{\mathrm{load}},
\label{eq:wecc_total_power}
\end{equation}
where
\begin{equation}
F_{\mathrm{st}} = 1 - F_{\mathrm{ma}} - F_{\mathrm{mb}} - F_{\mathrm{mc}} - F_{\mathrm{md}} - F_{\mathrm{el}}.
\label{eq:static_fraction}
\end{equation}

Motors A, B, and C are three-phase induction machines representing compressors ($e_{\mathrm{trq}}=0$), fans ($e_{\mathrm{trq}}=2$), and pumps ($e_{\mathrm{trq}}=2$), respectively. Each is represented by a fifth-order electromechanical model\cite{Ma2020WECCMath}. Motor D is a performance-based model for single-phase air-conditioning compressors that captures stalling behavior using piecewise $P$--$V$ and $Q$--$V$ curves together with a thermal relay, contactor, and under-voltage relay\cite{Lesieutre2008ACMotor}.

\subsection{FIDVR Mechanism}
When a transmission fault depresses the load-bus voltage below the stall threshold ($V_{\mathrm{stall}}\approx 0.6$ pu), Motor D compressors stall and transition from their normal operating state to an effective stall impedance
\begin{equation}
Z_{\mathrm{stall}} = R_{\mathrm{stall}} + jX_{\mathrm{stall}}.
\label{eq:stall_impedance}
\end{equation}
The stalled motors draw significantly higher current than during normal operation, which prevents the voltage from returning quickly to nominal after fault clearance. Recovery then occurs gradually as the thermal relay integrates the heating effect of the stalled state through the thermal time constant $T_{\mathrm{TH}}$ and progressively trips the stalled fraction\cite{Kosterev2006WECC}.

\section{Proposed Data-Center Composite Load Model}

The proposed \name is connected to the WECC CLM load bus as an additional aggregate load component, as illustrated in Fig.~\ref{fig:architecture}. The model is parameterized by the rated data-center power $P_{\mathrm{dc}}$, which is varied from 100~MW to 500~MW in the study scenarios. The total nominal data-center demand is divided among IT, cooling, and auxiliary loads using the fractions $F_{\mathrm{it}}$, $F_{\mathrm{cool}}$, and $F_{\mathrm{aux}}$, respectively. The values summarized in Table~\ref{tab:system_params} are the parameters adopted for the present simulation study and should be interpreted as configurable scenario parameters rather than universal characteristics of data centers.

The IT component represents servers, accelerators, storage, and networking equipment supplied through an aggregate UPS representation. The cooling component is divided into motor-driven and VFD-driven portions so that stall-sensitive and ride-through behaviors can coexist within the same facility model. The auxiliary component represents support demand that is not modeled separately.

The proposed structure is intended for transmission-level dynamic studies in which representing individual servers, power converters, and cooling devices in detail is neither necessary nor computationally practical. For application to a specific facility, the load fractions, UPS thresholds, cooling-system parameters, and workload profiles should be calibrated using site measurements or manufacturer-provided information. Such site-specific calibration and field validation are outside the scope of the present study.

\begin{table}[!t]
\caption{DC Composite Load Model System-Level Parameters}
\label{tab:system_params}
\centering
\footnotesize
\setlength{\tabcolsep}{3pt}
\renewcommand{\arraystretch}{1.1}
\begin{tabular}{>{\raggedright\arraybackslash}p{0.12\linewidth} >{\raggedright\arraybackslash}p{0.38\linewidth} >{\centering\arraybackslash}p{0.38\linewidth}}
\toprule
Parameter & Description & Value \\
\midrule
$P_{\mathrm{dc}}$ & DC rated power & \SIrange{100}{500}{MW} \\
$F_{\mathrm{it}}$ & IT load fraction & 0.60 \\
$F_{\mathrm{cool}}$ & Cooling load fraction & 0.30 \\
$F_{\mathrm{aux}}$ & Auxiliary load fraction & 0.10 \\
$F_{\mathrm{vfd}}$ & VFD share of cooling load & 0.40 \\
\bottomrule
\end{tabular}
\vspace{-10pt}
\end{table}

\begin{table}[!t]
\caption{UPS and Cooling-Load Parameters}
\label{tab:ups_cooling}
\centering
\footnotesize
\setlength{\tabcolsep}{3pt}
\renewcommand{\arraystretch}{1.1}
\begin{tabular}{>{\raggedright\arraybackslash}p{0.24\linewidth} >{\raggedright\arraybackslash}p{0.48\linewidth} >{\centering\arraybackslash}p{0.18\linewidth}}
\toprule
Parameter & Description & Value \\
\midrule
\multicolumn{3}{l}{\textit{UPS parameters}} \\
$V_{\mathrm{ups,trip}}$ & Battery transfer threshold (pu) & 0.90 \\
$V_{\mathrm{ups,return}}$ & Grid return threshold (pu) & 0.95 \\
$T_{\mathrm{ups,batt}}$ & Battery hold-up time & \SI{30}{s} \\
$T_{\mathrm{diesel}}$ & Diesel start time & \SI{12}{s} \\
$T_{\mathrm{ups,xfer}}$ & Transfer-switch time & \SI{0.004}{s} \\
$\mathrm{pf}_{\mathrm{it}}$ & IT load power factor & 0.95 \\
\midrule
\multicolumn{3}{l}{\textit{Cooling parameters (motor-driven portion)}} \\
$R_{\mathrm{stall}}$ & Stall resistance (pu) & 0.10 \\
$X_{\mathrm{stall}}$ & Stall reactance (pu) & 0.12 \\
$V_{\mathrm{stall}}$ & Stall voltage (pu) & 0.55 \\
$T_{\mathrm{stall}}$ & Stall delay & \SI{0.05}{s} \\
$V_{\mathrm{rst}}$ & Restart voltage (pu) & 0.90 \\
$T_{\mathrm{rst}}$ & Restart delay & \SI{2.0}{s} \\
$T_{\mathrm{TH}}$ & Thermal time constant & \SI{60}{s} \\
$\theta_{\mathrm{th},1}/\theta_{\mathrm{th},2}$ & Thermal trip levels (pu) & 1.6 / 2.8 \\
$\mathrm{pf}_{\mathrm{cool}}$ & Cooling-load power factor & 0.85 \\
\bottomrule
\end{tabular}
\vspace{-10pt}
\end{table}

\subsection{IT Load and UPS Supervisory Control}

The IT component represents the aggregate demand of servers, accelerators, storage, and networking equipment supplied through a UPS-based power-conditioning system. At the transmission-system level considered in this work, the supervisory-control problem is to determine whether the IT demand is visible to the utility grid during and after a voltage disturbance. The control is implemented as a deterministic rule-based state machine rather than as an optimization-based controller. This aggregate representation is motivated by the source-transition structure considered in our related data-center supervisory-coordination work~\cite{Ahmed2026ModalSupervisory}, while the present model simplifies those operational transitions into grid, battery, and diesel states suitable for integration with the WECC CLM.

The grid-side active power of the IT component is modeled as
\begin{equation}
P_{\mathrm{it}}(t)
=
u_{\mathrm{g}}(t)F_{\mathrm{it}}P_{\mathrm{dc}}P_{\mathrm{ai}}(t),
\end{equation}
where $F_{\mathrm{it}}$ is the IT-load fraction, $P_{\mathrm{dc}}$ is the rated data-center power, $P_{\mathrm{ai}}(t)$ is the normalized workload signal, and $u_{\mathrm{g}}(t)$ represents the grid-connection status of the IT load. The corresponding reactive power is calculated from the assumed IT-load power factor.

The UPS supervisory logic contains three operating states: grid, battery, and diesel. In grid mode, $u_{\mathrm{g}}=1$, and the IT demand is supplied from the utility grid. If the measured load-bus voltage falls below $V_{\mathrm{ups,trip}}$ and the condition persists for the transfer delay $T_{\mathrm{ups,xfer}}$, the state changes from grid to battery and $u_{\mathrm{g}}$ becomes zero. The IT demand is therefore removed from the grid-side power command. If the low-voltage condition persists beyond $T_{\mathrm{diesel}}$, the supervisory state changes from battery to diesel, while the IT demand remains isolated from the utility grid in the present representation. When the bus voltage recovers above $V_{\mathrm{ups,return}}$, the state returns to grid mode and the workload-dependent IT demand is restored.

The difference between $V_{\mathrm{ups,trip}}$ and $V_{\mathrm{ups,return}}$ provides hysteresis between the transfer and return conditions. Because the restored grid-side demand depends on the instantaneous value of $P_{\mathrm{ai}}(t)$, reconnection can introduce a workload-dependent active-power step that interacts with the recovering composite load.

The UPS representation is intentionally simplified for aggregate phasor-domain stability studies. It captures supervisory source transfer and the resulting grid-side visibility of the IT load, but it does not resolve converter switching, DC-link dynamics, battery state-of-charge evolution, converter current limiting, synchronization controls, detailed transfer-switch transients, or diesel-generator electromechanical dynamics. Accordingly, $T_{\mathrm{ups,xfer}}$ represents an event delay in the aggregate supervisory logic rather than a switching-level UPS transfer transient.

\subsection{Cooling Load}

The cooling system is modeled as a combination of motor-driven and VFD-driven components, weighted by $(1-F_{\mathrm{vfd}})$ and $F_{\mathrm{vfd}}$, respectively. This structure reflects the mixed composition of modern data-center cooling plants. The cooling-load parameters used in the model, including stall voltage, restart delay, thermal time constant, thermal trip thresholds, and power factor, are summarized in Table~\ref{tab:ups_cooling}.

The motor-driven portion follows a stall-and-recovery representation similar to the WECC Motor~D model, but with parameters chosen to reflect the slower restart and higher thermal inertia of industrial cooling equipment \cite{Lesieutre2008ACMotor,GinzburgGanz2026Survey,Sun2022DCEmulator}. Its stall-state active and reactive powers are given by
\begin{equation}
P_{\mathrm{cool,stall}} = G_{\mathrm{stall}}V^2,
\qquad
Q_{\mathrm{cool,stall}} = -B_{\mathrm{stall}}V^2,
\end{equation}
where
\begin{equation}
G_{\mathrm{stall}}=\frac{R_{\mathrm{stall}}}{R_{\mathrm{stall}}^2+X_{\mathrm{stall}}^2},
\qquad
B_{\mathrm{stall}}=\frac{X_{\mathrm{stall}}}{R_{\mathrm{stall}}^2+X_{\mathrm{stall}}^2}.
\end{equation}
The thermal state evolves as
\begin{equation}
\frac{d\theta}{dt}=\frac{I^2R_{\mathrm{stall}}-\theta}{T_{\mathrm{TH}}},
\end{equation}
and the stalled fraction is tripped progressively between $\theta_{\mathrm{th,1}}$ and $\theta_{\mathrm{th,2}}$.

The VFD-driven portion is assumed to maintain approximately constant power down to a minimum ride-through voltage, below which its power decreases linearly. In addition, the total cooling demand is coupled to the IT workload through
\begin{equation}
f_{\mathrm{cool}} = 0.6 + 0.4\,P_{\mathrm{ai}}(t),
\end{equation}
so that higher workload produces higher cooling demand.

The cooling-load parameters in Table~\ref{tab:ups_cooling} define the simulation scenarios considered in this study and have not been calibrated against measurements from a particular data center. Similarly, the VFD-driven portion is represented through an aggregate voltage-dependent power characteristic rather than a detailed converter-control model. This level of representation is intended for transmission-level voltage-recovery studies and does not capture converter switching, DC-link dynamics, harmonic effects, or equipment-specific protection. For facility-specific studies, these parameters should be replaced or calibrated using measured data or manufacturer-provided characteristics.

\subsection{AI Workload Profile Generator}

\begin{figure}[!t]
    \centering
    \includegraphics[width=\linewidth]{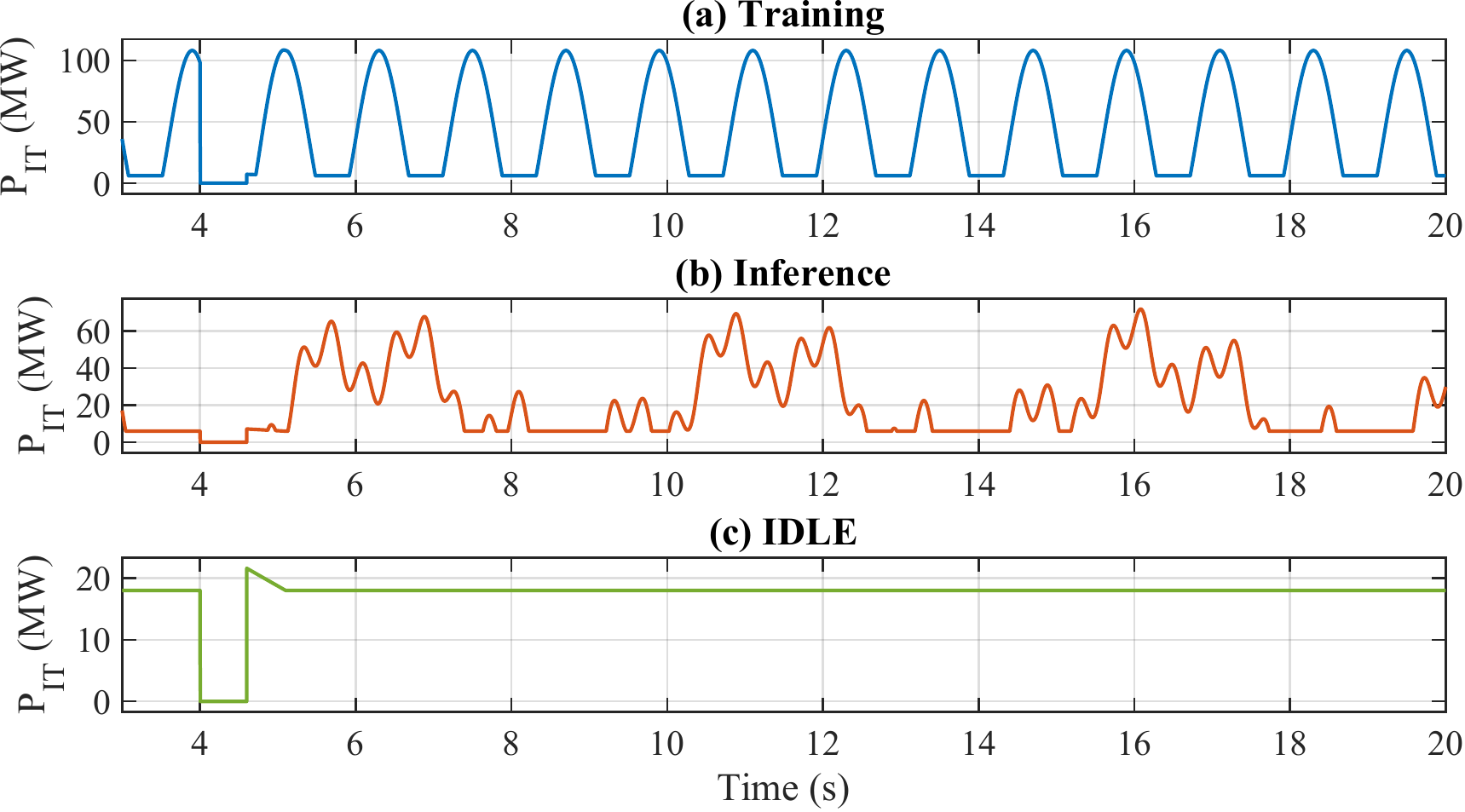}
    \caption{Representative synthetic IT-load power profiles used for the three AI workload modes in the proposed data-center composite load model: (a) training (b) inference (c) idle.
    \vspace{-10pt}}
\label{fig:IT_Load}
\end{figure}

\begin{table}[!t]
\caption{AI Workload Profile Parameters}
\label{tab:workload}
\centering
\footnotesize
\setlength{\tabcolsep}{3pt}
\renewcommand{\arraystretch}{1.1}
\begin{tabular}{>{\raggedright\arraybackslash}p{0.32\linewidth} >{\raggedright\arraybackslash}p{0.40\linewidth} >{\centering\arraybackslash}p{0.18\linewidth}}
\toprule
Parameter & Description & Value \\
\midrule
\multicolumn{3}{l}{\textit{Training mode}} \\
$P_{\mathrm{base}}$ & Base GPU utilization & 0.30 pu \\
$T_{\mathrm{ckpt}}/T_{\mathrm{ckpt,dur}}$ & Checkpoint interval / duration & 60 s / 2 s \\
$P_{\mathrm{ckpt}}$ & Power during checkpoint & 0.40 pu \\
$A_{\mathrm{comm}}/T_{\mathrm{comm}}$ & Sync-oscillation amplitude / period & 0.60 / 1.2 \\
\midrule
\multicolumn{3}{l}{\textit{Inference mode}} \\
$P_{\mathrm{base}}$ & Base utilization & 0.15 pu \\
$A_1/T_1,\;A_2/T_2,\;A_3/T_3$ & Multi-frequency burst parameters & 0.25/5, 0.15/1.3, 0.10/0.4 \\
\midrule
\multicolumn{3}{l}{\textit{Idle mode}} \\
$P_{\mathrm{idle}}$ & Idle utilization & 0.15 pu \\
\bottomrule
\end{tabular}
\vspace{-10pt}
\end{table}

The normalized IT demand $P_{\mathrm{ai}}(t)$ is represented using three synthetic operating profiles: training, inference, and idle. Training combines a 0.30~pu base demand with periodic communication-related variation and checkpoint events. Inference uses a 0.15~pu base demand with multiple frequency components to represent bursty activity, while idle remains constant at 0.15~pu. The resulting profiles are limited to $0.05 \leq P_{\mathrm{ai}}(t) \leq 1.0$~pu and are shown in Fig.~\ref{fig:IT_Load}.

These profiles are deterministic study inputs rather than measured traces from a specific facility. They are used to compare how workload level and variability affect UPS-mediated load removal and restoration. Their qualitative distinction is motivated by previously reported differences between training and inference power behavior~\cite{Patel2024LLM}.

\section{Simulink PSPD Implementation}

\name is implemented in MATLAB/Simulink R2023b using Simscape Electrical~\cite{MathWorks2023Simscape}. The \texttt{powergui} block is configured for 60~Hz positive-sequence phasor simulation. The study system includes a 230~kV source with a 10~GVA short-circuit level, a 230/33~kV, 800~MVA transformer with $x_{\mathrm{xf}}=0.08$~pu, a feeder equivalent, and a 100~MW WECC composite load. Motors A--C are represented using per-unit asynchronous-machine blocks.

The data-center component is implemented using a Three-Phase Dynamic Load block driven by active- and reactive-power commands from a MATLAB Function block containing the UPS supervisory logic, cooling dynamics, and workload generator. A Unit Delay with $T_s=1$~ms breaks the voltage--power algebraic loop, and the load filtering time constant is 0.1~s. Although implemented in Simulink, the formulation can be transferred to other phasor-domain platforms supporting controlled load injections and event-driven state transitions. The model captures aggregate voltage-recovery dynamics but does not represent converter switching, phase imbalance, DC-link dynamics, or detailed protection interactions; these effects require an EMT implementation.

\begin{figure}[!t]
    \centering
    \includegraphics[width=\columnwidth]{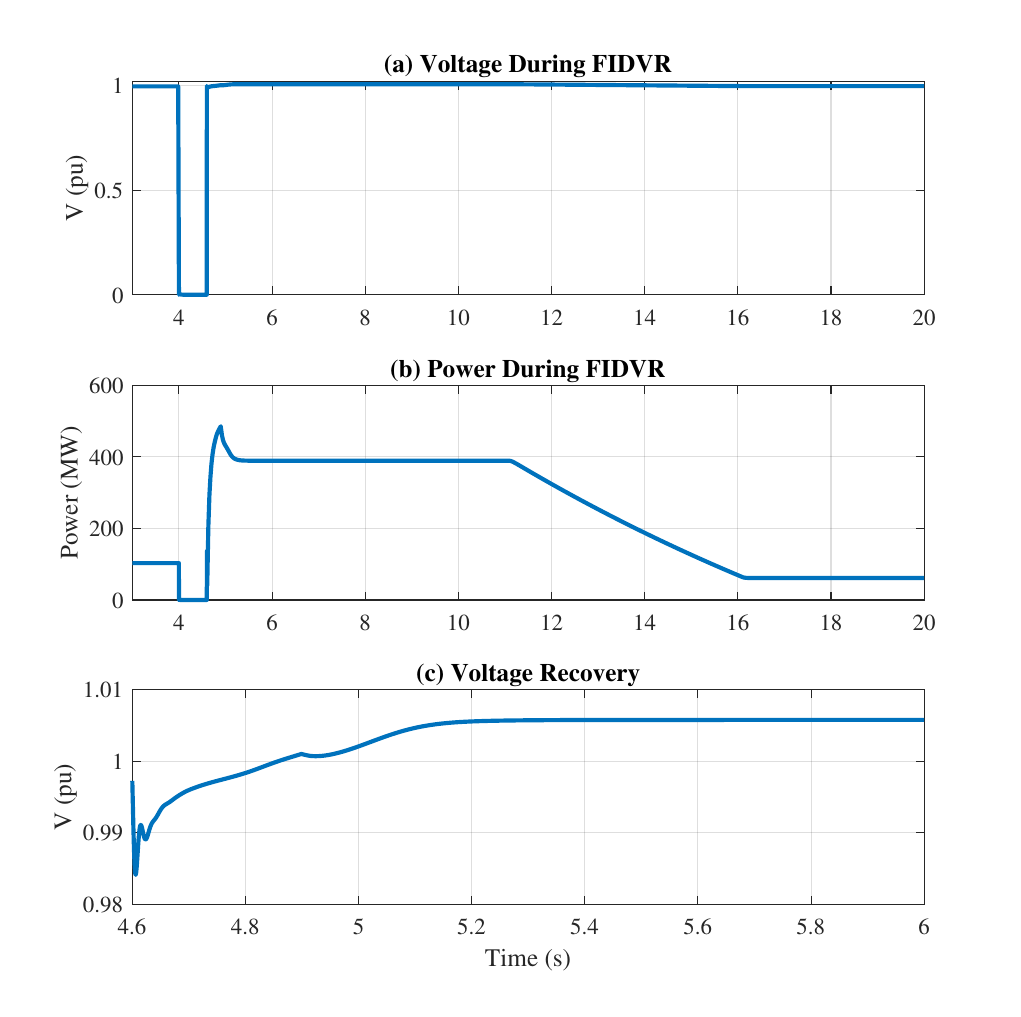}
    \caption{Reference FIDVR response. (a) Load-bus voltage during the disturbance and recovery period. (b) Total active power during the same interval. (c) Zoomed voltage recovery immediately after fault clearing, showing the short-timescale oscillatory transient.}
    \vspace{-7pt}
    \label{fig:reference_fidvr}
\end{figure}
\section{Case Studies}

\subsection{Baseline FIDVR Response}

Fig.~\ref{fig:reference_fidvr} shows the baseline FIDVR response of the reference system. The load-bus voltage remains close to nominal before the disturbance, then collapses sharply when the fault is applied at $t=4.0$~s. After fault clearing at $t=4.6$~s, the voltage does not recover instantaneously; instead, it exhibits the characteristic delayed recovery associated with FIDVR behavior. The total active power also drops during the fault and then rebounds rapidly after clearing, followed by a slower decline as the post-fault system response settles.

The zoomed view in Fig.~\ref{fig:reference_fidvr}(c) highlights the short-timescale oscillatory behavior immediately after fault clearing. A brief undershoot is followed by damped oscillations as the voltage returns toward its post-fault steady level. This reference response provides a useful baseline for comparing how the proposed \name modifies both the long-timescale recovery trajectory and the short-timescale oscillatory transient under different workload conditions.

\subsection{FIDVR Response with 200 MW Data-Center Load}

\begin{figure}[!t]
    \centering
    \includegraphics[width=\linewidth]{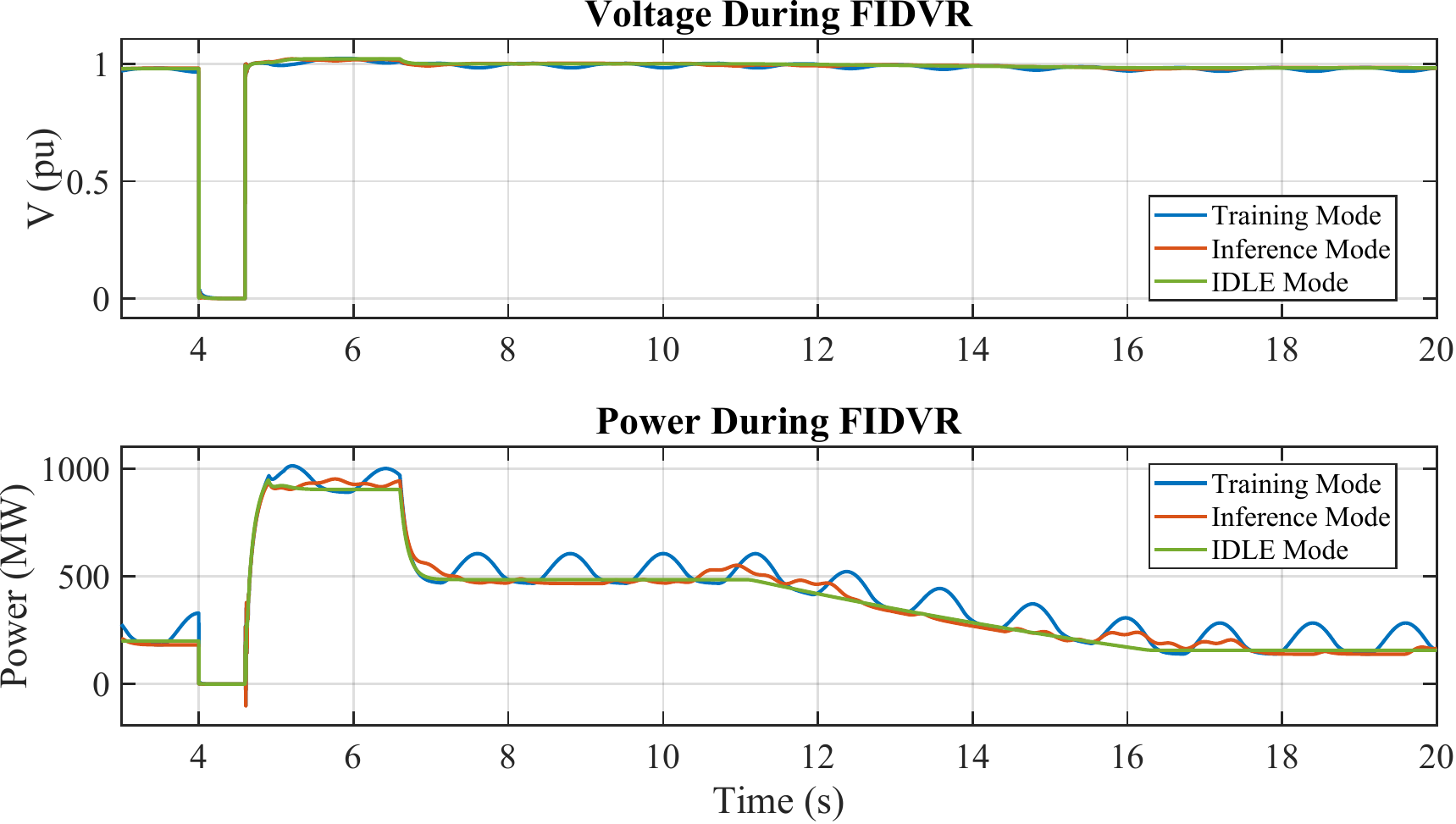}
    \caption{FIDVR response of the study system containing a 100~MW conventional WECC composite load and a 200~MW \name component under training, inference, and idle workload conditions. The fault is applied at $t=4.0$~s and cleared at $t=4.6$~s.}
    \vspace{-7pt}
    \label{fig:fidvr_main}
\end{figure}

Fig.~\ref{fig:fidvr_main} shows the response of the 100~MW WECC composite load combined with a 200~MW \name component under training, inference, and idle workloads. A three-phase fault is applied at $t=4.0$~s and cleared at $t=4.6$~s. During the fault, the UPS supervisory logic removes the IT demand from the grid, while the bus voltage collapses.

After fault clearance, the conventional Motor D load remains stalled, causing delayed voltage recovery governed mainly by its thermal-relay dynamics. Consequently, the long-duration voltage trajectories are similar across the three workloads, whereas the active-power responses differ because the restored IT demand depends on the workload level.

\subsection{AI Workload Mode Comparison}

Three simulations are performed with $P_{\mathrm{dc}}=200$~MW under training, inference, and idle workloads. As shown in Fig.~\ref{fig:fidvr_main}, the long-duration voltage-recovery trajectories are nearly identical because the UPS temporarily removes the IT demand from the grid, leaving the conventional stalled Motor D load to govern the overall FIDVR duration.

\begin{figure}[!t]
    \centering
    \includegraphics[width=\linewidth]{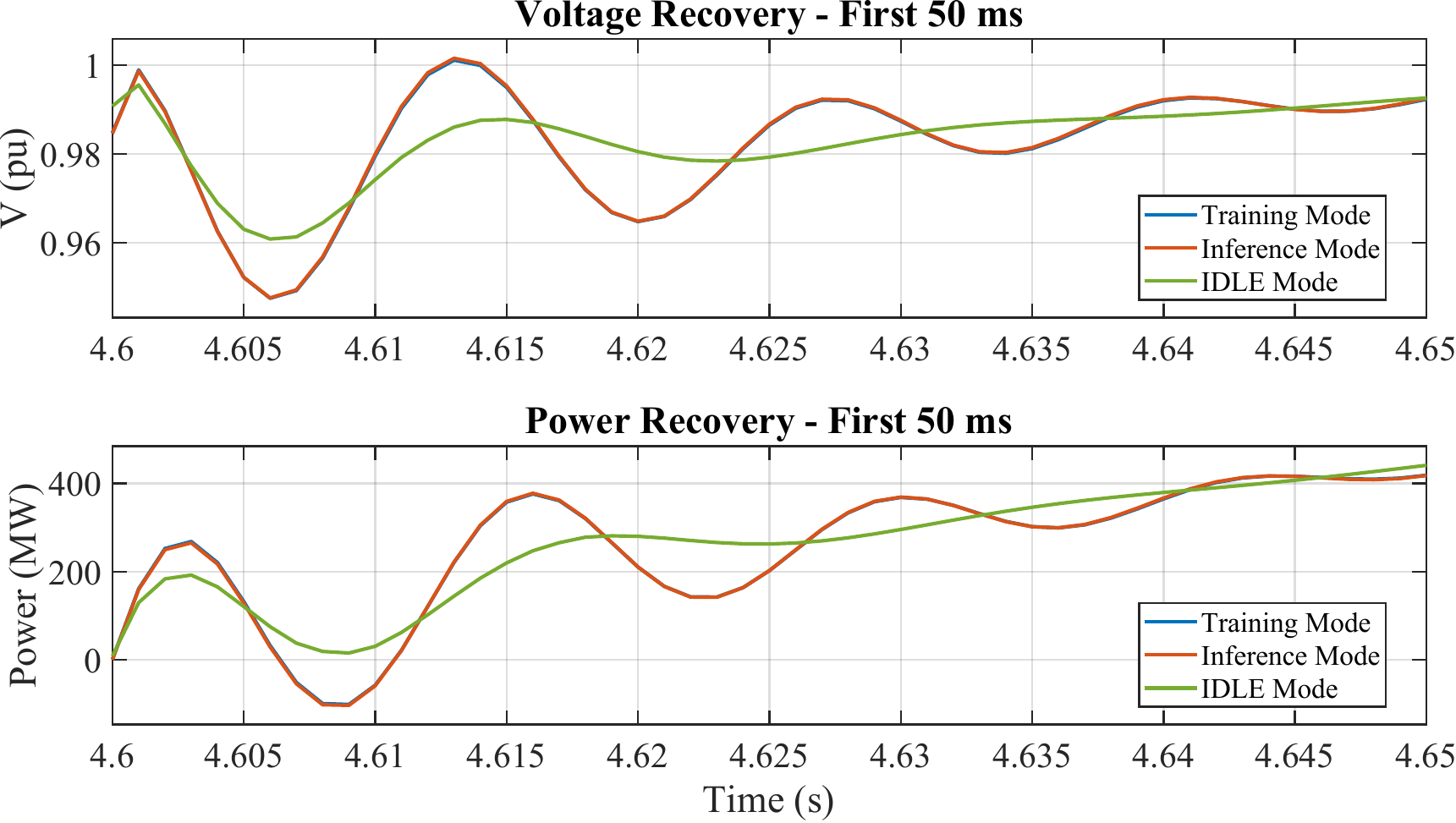}
    \caption{Aggregate post-fault voltage and active-power responses during the first 50~ms after fault clearance. Training and inference produce larger overall excursions than idle operation under the modeled workload and UPS restoration conditions.}
    \vspace{-7pt}
    \label{fig:postfault_zoom}
\end{figure}

\begin{figure}[!t]
    \centering
    \includegraphics[width=\columnwidth]{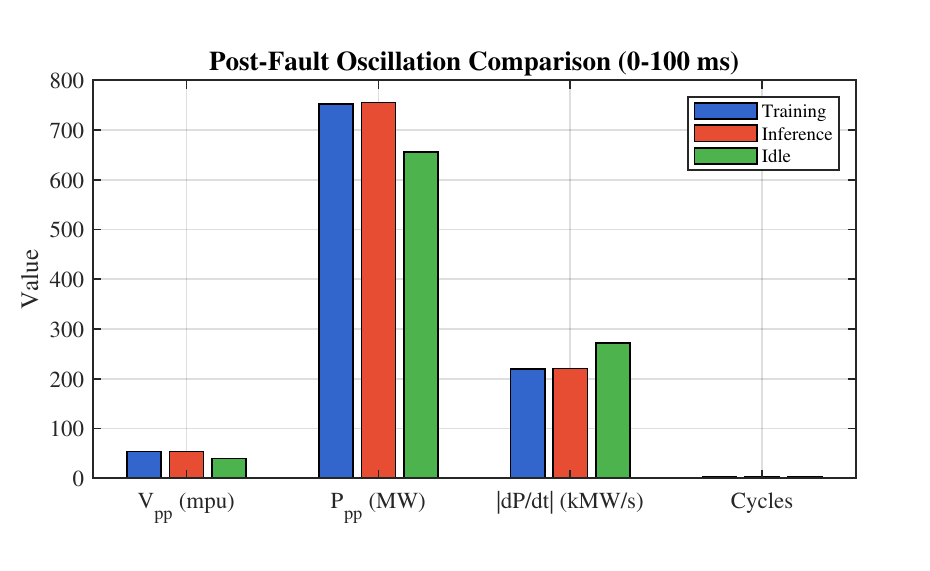}
    \vspace{-25pt}
    \caption{Post-fault metrics calculated over the first 100~ms after fault clearance. $V_{\mathrm{pp}}$ is reported in milli-per-unit (mpu), $P_{\mathrm{pp}}$ in MW, and $\max|dP/dt|$ in kMW/s.}
    \vspace{-7pt}
    \label{fig:bar_metrics}
\end{figure}

The workload effect is more evident immediately after fault clearance. Fig.~\ref{fig:postfault_zoom} shows larger short-timescale voltage and active-power excursions for training and inference than for idle operation. The first-100-ms metrics in Fig.~\ref{fig:bar_metrics} show voltage variations of approximately 50--52~mpu and active-power swings of 751--755~MW for training and inference, compared with approximately 36~mpu and 654~MW for idle. Although idle exhibits the largest isolated $\max|dP/dt|$ at approximately 269~kMW/s, its overall voltage and power excursions remain smaller. Thus, workload type primarily affects the short-timescale post-fault response rather than the overall FIDVR duration.

\subsection{Comparison with Conventional Load Models}

Table~\ref{tab:comparison} compares the proposed \name with the conventional WECC Motor D component. Motor D represents aggregated single-phase air-conditioning compressor dynamics, including voltage-induced stalling and delayed thermal recovery~\cite{Lesieutre2008ACMotor,Kosterev2008WECCProgress}. In contrast, \name is an additional aggregate component for data-center loads and includes UPS-supported IT demand, mixed motor/VFD cooling, auxiliary demand, and workload-dependent behavior.

The main distinction is structural. In \name, the IT load can be temporarily isolated from the grid through UPS supervisory logic and restored after voltage recovery, producing a workload-dependent rebound that is not represented by Motor D. The mixed cooling model further captures the coexistence of stall-sensitive and VFD-driven loads. The comparison is therefore intended to highlight additional data-center-specific dynamics rather than provide field validation. The numerical parameters used in this study are configurable simulation values and should be calibrated with facility-specific data.

\begin{table}[!t]
\caption{Feature-Level Comparison of WECC Motor D and Proposed \name}
\label{tab:comparison}
\centering
\footnotesize
\setlength{\tabcolsep}{3pt}
\renewcommand{\arraystretch}{1.1}
\begin{tabular}{>{\raggedright\arraybackslash}p{0.36\linewidth} c c}
\toprule
Characteristic & Motor D & \name \\
\midrule
Application & Residential A/C & AI data center \\
$V_{\mathrm{stall}}$ (pu) & 0.60 & 0.55 \\
$T_{\mathrm{rst}}$ (s) & 0.30 & 2.0 \\
$T_{\mathrm{TH}}$ (s) & 20 & 60 \\
VFD portion & None & 40\% \\
UPS ride-through & No & Yes \\
Load rebound & No & Yes \\
Workload variability & None & AI-dependent \\
Grid-side IT demand after UPS transfer & N/A & Zero \\
\bottomrule
\end{tabular}
\vspace{-10pt}
\end{table}

\section{Conclusion}

This paper presented \name, a workload-aware extension of the WECC composite load model incorporating UPS-supported IT demand, mixed motor/VFD cooling, auxiliary load, and AI workload profiles. For the studied system, long-duration FIDVR recovery was governed mainly by the conventional stalled Motor D load, while training and inference produced larger short-timescale voltage and active-power excursions than idle operation. The current model uses configurable parameters and aggregate phasor-domain representations. Future work will pursue measurement-based calibration, field validation, and detailed EMT modeling, with high-resolution monitoring frameworks such as GridStream~\cite{GridStream2026} supporting disturbance-data collection.The \name implementation and example study files are publicly available at \url{https://github.com/KIBRIA-SAROARE/Extended-WECC-Composite-Load-Model-with-DC} to support reproduction of the reported cases, parameter modification, and comparison with other load-modeling approaches.

\medskip
\noindent \textbf{Acknowledgments.}
This research was funded in part by the U.S. National Science Foundation under Grant Number OIA-2437963 and the Louisiana Board of Regents; as well as by a gift from the COES at Louisiana Tech University. Any opinions, findings, conclusions or recommendations expressed in this document are those of the authors and do not necessarily reflect the views or official policies, expressed or implied, of the sponsoring organizations.

\bibliographystyle{IEEEtran}
\bibliography{refs}

\end{document}